\documentclass{ws-procs9x6}

\def\etal {{\it et al.}}
\def\sb{\overline{s}}

\begin{document}

\title{CONSTRAINTS ON VIOLATIONS OF LORENTZ SYMMETRY FROM GRAVITY PROBE B}

\author{JAMES M.\ OVERDUIN$^*$ and RYAN D.\ EVERETT}

\address{Department of Physics, Astronomy \& Geosciences, Towson University\\
Towson, MD 21252, U.S.A.\\
$^*$E-mail: joverduin@towson.edu}

\author{QUENTIN G.\ BAILEY}

\address{Department of Physics, Embry-Riddle Aeronautical University\\
Prescott, AZ 86301, U.S.A.\\
E-mail: baileyq@erau.edu}

\begin{abstract}
We use the final results from Gravity Probe~B to set new upper limits on the gravitational sector of the Standard-Model Extension, including for the first time the coefficient associated with the time-time component of the new field responsible for inducing local Lorentz violation in the theory.
\end{abstract}

\bodymatter

\section*{ }

The minimal pure-gravity sector of the Standard-Model Extension (SME) is characterized by nine independent coefficients $\bar{s}^{\text{\tiny AB}}$ corresponding to the vacuum expectation values of a new tensor field whose couplings to the traceless part of the Ricci tensor induce spontaneous violations of local Lorentz symmetry \cite{K04}.
These coefficients are assumed to be constant in the asymptotically flat (Minkowski) limit. Most are constrained either individually or in various combinations by existing experiments and observations \cite{AR11}, but no limits have yet been placed on the $\bar{s}^{\text{\tiny TT}}$ coefficient.

Gravity Probe~B (GPB) was a satellite experiment launched in 2004 to measure the geodetic and frame-dragging effects predicted by General Relativity (GR). As shown by Bailey and Kostelecky in 2006 \cite{BK06}, the orientation of a gyroscope in orbit around a spinning central mass like the earth is sensitive to seven of the nine $\bar{s}^{\text{\tiny AB}}$ coefficients, {\em including\/} $\bar{s}^{\text{\tiny TT}}$. Following earlier preliminary work \cite{O07}, our goal here is to calculate the resulting constraints using the recently released final results from GPB \cite{E11}.

\begin{figure}[t!]
\begin{center}
\psfig{file=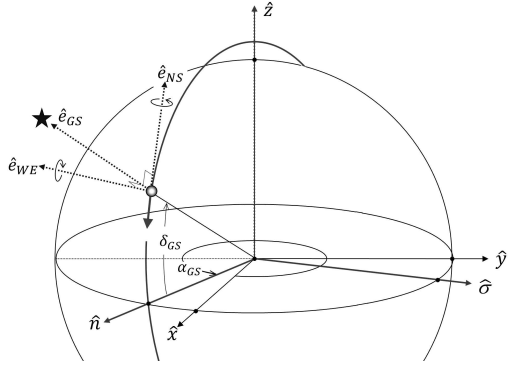,width=2.62in}
\end{center}
\caption{Experimental results are expressed in GPB coordinates $(\hat{e}_{\text{\tiny GS}},\hat{e}_{\text{\tiny NS}},\hat{e}_{\text{\tiny WE}})$. Theoretical SME predictions are derived in the $(\hat{n},\hat{\sigma},\hat{z})$ system. Both are ultimately referred to Sun-centered inertial coordinates $(\hat{x},\hat{y},\hat{z})$, where $\hat{x}$ points toward the vernal equinox.}
\label{fig:coords}
\end{figure}

Within GR the geodetic and frame-dragging precession rates of a gyroscope with position $\vec{r}$ and velocity $\vec{v}$ in orbit around a central mass $M$ with moment of inertia $I$ and angular velocity $\vec{\omega}$ are:
\begin{equation}
\vec{\Omega}_{\text{g,\tiny{GR}}} = \left(\frac{3}{2}\frac{GM}{c^2r^3}\right)\vec{r}\times\vec{v} \label{GRprecessions} \;\;\; , \;\;\;
\vec{\Omega}_{\text{fd,\tiny{GR}}} = \frac{GI}{c^2r^3}\left[\frac{3\vec{r}}{r^2}(\vec{\omega}\cdot\vec{r})-\vec{\omega}\right] .
\end{equation}
The combined precession $\vec{\Omega}_{\text{\tiny GR}}=\vec{\Omega}_{\text{g,\tiny{GR}}}+\vec{\Omega}_{\text{fd,\tiny{GR}}}$ causes the unit spin vector $\hat{S}$ of the gyroscope to undergo a relativistic drift
$\vec{R} \equiv d\hat{S}/dt=\vec{\Omega}_{\text{\tiny GR}}\times\hat{S}$.
Averaging over a circular, polar orbit of radius $r_0$ around a spherically symmetric central mass, one obtains
\begin{equation}
\vec{R}_{\text{g,\tiny{GR}}} = -\frac{3(GM)}{2\,c^2r_0^{5/2}}^{3/2} \!\!\!\!\! \hat{e}_{\text{\tiny NS}} \;\;\; , \;\;\;
\vec{R}_{\text{fd,\tiny{GR}}} = -\frac{GI\omega\cos\delta_{\text{\tiny GS}}}{2\,c^{2}r_0^3} \, \hat{e}_{\text{\tiny WE}} ,
\label{GRgeodeticDrift} \\
\end{equation}
where $\hat{e}_{\text{\tiny GS}}$ points toward the guide star (located in the orbit plane at right ascension $\alpha_{\text{\tiny GS}}$ and declination $\delta_{\text{\tiny GS}}$), $\hat{e}_{\text{\tiny WE}}$ is an orbit normal pointing along the cross-product of $\hat{e}_{\text{\tiny GS}}$ and the unit vector $\hat{z}$ (aligned with the earth's rotation axis) and $\hat{e}_{\text{\tiny NS}}$ is a tangent to the orbit directed along $\hat{e}_{\text{\tiny WE}}\times\hat{e}_{\text{\tiny GS}}$ (Fig.~\ref{fig:coords}).
The choice of polar orbit orthogonalizes the two effects so that $\vec{R}_{\text{g,\tiny{GR}}}$ points entirely along $\hat{e}_{\text{\tiny NS}}$ and $\vec{R}_{\text{fd,\tiny{GR}}}$ points entirely along $\hat{e}_{\text{\tiny WE}}$. 

For GPB with guide star IM~Pegasi, $r_0=7018.0$~km, $\delta_{\text{\tiny GS}}=16.841^{\circ}$, $R_{\text{g,\tiny GR}}=6606.1$~mas/yr (including oblateness) and $R_{\text{fd,\tiny GR}}=39.2$~mas/yr where mas=milliarcsecond. The final joint results for all four gyros indicate that $R_{\text{{\tiny NS},obs}}=6601.8\pm18.3$~mas/yr and $R_{\text{{\tiny WE},obs}}=37.2\pm7.2$~mas/yr with 1$\sigma$ uncertainties \cite{E11}. Thus the NS and WE components of relativistic drift rate may deviate from the predictions of GR by at most $\Delta R_{\text{\tiny NS}} < |R_{\text{g,\tiny GR}}-R_{\text{{\tiny NS},obs}}|$ = 22.6~mas/yr and  $\Delta R_{\text{\tiny WE}} < |R_{\text{fd,\tiny GR}}-R_{\text{{\tiny WE},obs}}|$ = 9.2~mas/yr.

Within the SME, Lorentz-violating terms introduce an additional ``anomalous'' relativistic drift $\Delta\vec{R}$ whose components along $\hat{n},\hat{\sigma}$ and $\hat{z}$ are given by Eqs.~(158-160) of Ref.~\refcite{BK06}.
Here $\hat{n}\equiv\hat{\sigma}\times\hat{z}$ and $\hat{\sigma}=-\hat{e}_{\text{\tiny WE}}$ is an orbit normal (Fig.~\ref{fig:coords}).
These equations may be expressed in the form
\begin{equation}
\Delta\vec{R} = \begin{pmatrix}
&\tfrac{1}{2}\,\omega_{\text{\tiny GS}}(\sb^{\text{\tiny YY}}-\sb^{\text{\tiny XX}})\sin{2\alpha_{\text{\tiny GS}}} + \omega_{\text{\tiny GS}}\sb^{\text{\tiny XY}}\cos{2\alpha_{\text{\tiny GS}}} & \\
&\omega_{\text{\tiny T}}\sb^{\text{\tiny TT}} + \omega_{\text{\tiny NS}}(\sb^{\text{\tiny XX}}\sin^2{\alpha_{\text{\tiny GS}}} - \sb^{\text{\tiny XY}}\sin{2\alpha_{\text{\tiny GS}}} + \sb^{\text{\tiny YY}}\cos^2{\alpha_{\text{\tiny GS}}}) \label{BKconstraints} & \\ 
&\omega_{\text{\tiny WE}}(\sb^{\text{\tiny YZ}}\cos{\alpha_{\text{\tiny GS}}} - \sb^{\text{\tiny XZ}}\sin{\alpha_{\text{\tiny GS}}}) & \\
\end{pmatrix} ,
\end{equation}
where $
\omega_{\text{\tiny GS}}=\omega_{\text{\tiny WE}}=\tfrac{5}{6}(1-3I/5Mr_0^2)\,R_{\text{g,\tiny GR}}=4603~\mbox{mas/yr, } 
\omega_{\text{\tiny T}}=\tfrac{3}{4}(1-I/3Mr_0^2)\,R_{\text{g,\tiny GR}}=4503~\mbox{mas/yr, }
\omega_{\text{\tiny NS}}=\tfrac{1}{12}(1+9I/Mr_0^2)\,R_{\text{g,\tiny GR}}=1904~\mbox{mas/yr}$
and $\alpha_{\text{\tiny GS}}=343.26^{\circ}$.
To transform to GPB coordinates, we reflect across the orbit plane and rotate about $\hat{\sigma}$ by $\delta_{\text{\tiny GS}}$.
The resulting drift rates along the GS, NS and WE axes are
\begin{equation}
\Delta\vec{R}=\left(\begin{smallmatrix}
\begin{aligned}
& \omega_{\text{\tiny GS}}\!\left[\tfrac{1}{2}(\sb^{\text{\tiny YY}}-\sb^{\text{\tiny XX}})\sin{2\alpha_{\text{\tiny GS}}}\cos{\delta_{\text{\tiny GS}}}+\sb^{\text{\tiny XY}}\cos{2\alpha_{\text{\tiny GS}}}\cos{\delta_{\text{\tiny GS}}}\right.\\
&\hspace{18mm}-\left.\sb^{\text{\tiny XZ}}\sin{\alpha_{\text{\tiny GS}}}\sin{\delta_{\text{\tiny GS}}}+\sb^{\text{\tiny YZ}}\cos{\alpha_{\text{\tiny GS}}}\sin{\delta_{\text{\tiny GS}}}\right]\\
& -\omega_{\text{\tiny T}}\sb^{\text{\tiny TT}}-\omega_{\text{\tiny NS}}(\sb^{\text{\tiny XX}}\sin^2\alpha_{\text{\tiny GS}}-\sb^{\text{\tiny XY}}\sin{2\alpha_{\text{\tiny GS}}}+\sb^{\text{\tiny YY}}\cos^2\alpha_{\text{\tiny GS}})\\
& \omega_{\text{\tiny WE}}\!\left[\tfrac{1}{2}(\sb^{\text{\tiny XX}}-\sb^{\text{\tiny YY}})\sin{2\alpha_{\text{\tiny GS}}}\sin{\delta_{\text{\tiny GS}}}-\sb^{\text{\tiny XY}}\cos{2\alpha_{\text{\tiny GS}}}\sin{\delta_{\text{\tiny GS}}}\right.\\
&\hspace{18mm}-\left.\sb^{\text{\tiny XZ}}\sin\alpha_{\text{\tiny GS}}\cos\delta_{\text{\tiny GS}}+\sb^{\text{\tiny YZ}}\cos\alpha\cos\delta_{\text{\tiny GS}}\right]\\
\end{aligned}
\end{smallmatrix}\right) .
\vspace{-3mm}
\end{equation}
Numerically,
\begin{eqnarray}
\Delta R_{\text{\tiny GS}} & = &
1215 \sb^{\text{\tiny XX}} + 3674 \sb^{\text{\tiny XY}} + 384 \sb^{\text{\tiny XZ}} - 1215 \sb^{\text{\tiny YY}} + 1277 \sb^{\text{\tiny YZ}} , \nonumber \\
\Delta R_{\text{\tiny NS}} & = &
-4503 \sb^{\text{\tiny TT}} - 158 \sb^{\text{\tiny XX}} - 1050 \sb^{\text{\tiny XY}} - 1746 \sb^{\text{\tiny YY}} \label{numericalLimits} , \\
\Delta R_{\text{\tiny WE}} & = &
-368 \sb^{\text{\tiny XX}} - 1112 \sb^{\text{\tiny XY}} + 1269 \sb^{\text{\tiny XZ}} + 368 \sb^{\text{\tiny YY}} + 4219 \sb^{\text{\tiny YZ}} , \nonumber
\end{eqnarray}
where $\Delta R_{\text{\tiny NS}}<22.6$ and $\Delta R_{\text{\tiny WE}}<9.2$ from GPB (all units in mas/yr).
The SME can accommodate precessions greater than those predicted by GR, unlike other extensions of the standard model where Einstein's theory is a limiting case \cite{OEW13}.
GPB does not constrain the GS component, since the gyro spin axes point along this direction by design. The GS and WE components are linear combinations of $\sb^{\text{\tiny XY}},\sb^{\text{\tiny XZ}},\sb^{\text{\tiny YZ}}$ and $(\sb^{\text{\tiny XX}}-\sb^{\text{\tiny YY}})$, so they are superseded in any case by existing constraints, which read\cite{AR11,B07}:
\begin{eqnarray}
& & |\sb^{\text{\tiny XY}}| < (0.6 \pm 1.5) \times 10^{-9} \label{xyConstraint} \\
& & |\sb^{\text{\tiny XZ}}| < (2.7 \pm 1.4) \times 10^{-9} \label{xzConstraint} \\
& & |\sb^{\text{\tiny YZ}}| < (0.6 \pm 1.4) \times 10^{-9} \label{yzConstraint} \\
& & |\sb^{\text{\tiny XX}} - \sb^{\text{\tiny YY}}| < (1.2 \pm 1.6) \times 10^{-9} \label{xx-yyConstraint} \\
& & |\sb^{\text{\tiny XX}} + \sb^{\text{\tiny YY}} - 2\sb^{\text{\tiny ZZ}}| < (1.8 \pm 38)\phantom{.} \times 10^{-9} \label{xx-yy-2zzConstraint}
\end{eqnarray}
Thus in practice the only new GPB constraint on the SME comes from the NS component of Eqs.~(\ref{numericalLimits}), associated entirely with geodetic precession in standard GR. It reads:
\begin{equation}
|\sb^{\text{\tiny TT}}+0.035\sb^{\text{\tiny XX}}+0.23\sb^{\text{\tiny XY}}+0.39\sb^{\text{\tiny YY}}| < 5.0\times10^{-3} .
\label{gpbConstraint}
\end{equation}
To get seven conditions on seven unknowns, we supplement Eqs.~(\ref{xyConstraint}-\ref{gpbConstraint}) with the requirement that $\sb^{\text{\tiny AB}}$ be traceless, $|\sb^{\text{\tiny TT}} - \sb^{\text{\tiny XX}} - \sb^{\text{\tiny YY}} - \sb^{\text{\tiny ZZ}}| = 0$.\cite{BK06}
Inverting, we then find that
\begin{equation}
\sb^{\text{\tiny TT}} < 4.4 \times 10^{-3} \;\;\; , \;\;\;
\sb^{\text{\tiny XX}},\sb^{\text{\tiny YY}},\sb^{\text{\tiny ZZ}} < 1.5 \times 10^{-3} \nonumber .
\end{equation}
This constitutes the first experimental upper bound on $\sb^{\text{\tiny TT}}$. (Other tests such as light deflection are also sensitive to this coefficient at similar levels of precision \cite{TB11}.)
It also lifts a degeneracy between other existing limits, allowing us to extract individual upper bounds on $\sb^{\text{\tiny XX}},\sb^{\text{\tiny YY}}$ and $\sb^{\text{\tiny ZZ}}$.

One should also look at the effect of $\sb^{\text{\tiny AB}}$ on the equation of motion for the gyroscope\cite{BK06}. This has the effect of rescaling Newton's gravitational constant $G$, increasing our sensitivity to $\sb^{\text{\tiny TT}}$ and strengthening our limits by about 5\%\cite{BEO13}. If the actual orbit is not perfectly circular, as was the case for GPB (whose gyros remained in essentially perfect free fall around a non-spherically symmetric Earth), then additional $\sb^{\text{\tiny AB}}$-dependent terms are also introduced in the {\em leading-order\/} (GR) expressions for geodetic and frame-dragging precession, Eqs.~(\ref{GRprecessions}). These do not significantly alter the NS or geodetic constraint from GPB, but they do strengthen the WE or frame-dragging constraint so that it may potentially become competitive with existing limits. We will report on these results elsewhere.\cite{BEO13}


\end{document}